\documentclass[a4paper,11pt]{article}
\pdfoutput=1
\usepackage{jheppub}
\usepackage[T1]{fontenc}
\usepackage{lmodern}
\usepackage{bm}
\usepackage{mathtools}
\usepackage{booktabs}
\usepackage{microtype}
\usepackage{graphicx}
\usepackage{array}

\title{\boldmath Landau Theory of Black Brane Phase Transitions\\
from Kasner Singularities}

\author[a]{Hong-Da Lyu}

\affiliation[a]{Key Laboratory of Particle Physics and Particle Irradiation (MOE), \\Institute of Frontier and Interdisciplinary Science, \\Shandong University, Qingdao, Shandong 266237, China}

\emailAdd{hongdalyu@buaa.edu.cn}

\abstract{
We develop a Landau theory of a continuous black-brane phase
transition using the geometry of the black-hole interior.
The normal phase has a Reissner--Nordstr\"om Cauchy horizon,
whereas the scalarized phase ends at a Kasner singularity.
A conserved radial charge determines the grand potential
from either endpoint and permits a direct phase comparison.
At the critical point, we show analytically that the
quadratic correction to the conserved radial charge
vanishes. Using the inverse Kasner scalar velocity
as an interior order parameter, we then construct
a Landau potential whose coefficients are determined
entirely from the terminal singularity.
The Iyer--Wald first law relates this potential to
the conventional boundary Landau theory and fixes
its quartic coefficient.
For two models with different scalar self-interactions,
the resulting theory predicts the critical scaling
of the Kasner order parameter and the grand-potential
difference between the phases, in agreement with
fully backreacted numerical solutions.
}


\newcommand{\dd}{\mathrm d}
\newcommand{\calK}{\mathcal K}
\newcommand{\calQ}{\mathcal Q}

\newcommand{\calF}{\mathcal F}
\newcommand{\calV}{\mathcal V}

\newcommand{\calM}{\mathcal M}

\newcommand{\calH}{\mathcal H}
\newcommand{\eps}{\epsilon}

\newcommand{\RN}{\mathrm{RN}}

\newcommand{\K}{\mathrm K}

\begin{document}
\maketitle
\flushbottom

\section{Introduction}
\label{sec:introduction}

Scalar condensation can drive a continuous phase
transition from a Reissner--Nordstr\"om (RN) black brane
to a hairy black brane in asymptotically AdS spacetime
\cite{Gubser2008,HartnollHerzogHorowitz2008PRL,
HartnollHerzogHorowitz2008JHEP}.
While the transition is usually characterized by
the boundary condensate and the grand potential,
it also changes the geometry behind the event horizon.
In particular, scalar hair can destroy the RN Cauchy
horizon and replace it with a spacelike Kasner singularity
\cite{HartnollEtAl2020,CaiLiYang2021,
HartnollEtAl2021,AnLiYangYang2022}.

The approach to the singularity in hairy black holes
can involve a sequence of distinct dynamical regimes.
In AdS black holes, the interior evolution may feature
Einstein--Rosen bridge collapse, Josephson oscillations,
and transitions between successive Kasner epochs
\cite{HartnollEtAl2021,SwordVegh2022,CaiDuanLiYang2024,
Zhang:2025hkb}.
Related phenomena have also been found in helical
black holes \cite{LiuLyu2022Helical} and rotating
hairy black holes \cite{GaoLiuLyu2024,GaoLiuZhao2026}.
The asymptotic Kasner geometry has been studied
analytically in black holes with scalar or vector
hair \cite{Henneaux2022}.
In asymptotically flat Einstein--scalar black holes,
the Kasner exponents and near-singularity parameters
have also been related to the black-hole mass and
scalar charge \cite{XiongLu2026Kasner}.
More general anisotropic interiors may exhibit
BKL-type dynamics
\cite{DeClerckEtAl2024,CaceresEtAl2026BKL}.

The terminal Kasner geometry is also sensitive
to the phase structure of the boundary theory.
In holographic topological semimetals, the Kasner
exponents exhibit characteristic behavior across
topological phase transitions, remaining nearly
constant in the low-temperature topological phase
\cite{GaoLiuLyu2023}.
The Kasner exponents can change discontinuously
across first-order phase transitions
\cite{LiuLyuRaju2021,GaoLiuLyu2608},
while near continuous scalarization transitions
they exhibit critical scaling in both
asymptotically flat and AdS black holes
\cite{LiSunYang2026,SunEtAl2026,
GaoLiuLyu2608,GaoLiuLyu2609}.
In holographic models, this scaling can be
related to the scalar zero mode at the onset
of condensation, with different critical
behavior arising at multicritical points
\cite{GaoLiuLyu2608,GaoLiuLyu2609,LiuLyu2026}.

Beyond its critical behavior, the black hole
interior can leave signatures in thermal
correlation functions
\cite{AfkhamiJeddiEtAl2026,CeplakValach2026,Oh2026,ChakravartyZigdon2026}
and entanglement-based observables
\cite{HartmanMaldacena2013,CaceresShashiSun2024,
LiYang2026TEE}.
High-overtone quasinormal modes provide another
probe of the scaling behavior near the singularity,
with their asymptotic spectra encoding the Kasner
exponents and revealing transitions between
distinct scaling regimes
\cite{HartnollZhiboedov2026,XiaoNianLi2026,
CovielloEtAl2026a,CovielloEtAl2026b}.
Complementing these probes, an interior first law
relating horizon and Kasner singularity quantities
has been established \cite{XiongLu2608}
using the Iyer--Wald formalism
\cite{Wald1993,IyerWald1994}.

In this paper, we construct an interior Landau
potential for the neutral-scalarized RN--AdS$_4$
black branes of Ref.~\cite{GaoLiuLyu2609}.
We first show that a conserved radial charge
determines the grand potential from either the
RN Cauchy horizon or the terminal Kasner singularity.
At the critical point, the scalar zero-mode
equation leads to an exact cancellation of
the quadratic correction to this charge.
Using the reciprocal Kasner scalar velocity
$\eta_{\K}=4/v$ as an interior order parameter,
we then construct a local Landau potential
from the Kasner endpoint response.
The interior first law relates this potential
to the boundary Landau theory.
For two models with identical linear scalar
equations but different nonlinear interactions,
the resulting potential reproduces the critical
scaling and grand-potential difference obtained
from fully backreacted black brane solutions.
Its coefficients also agree with those extracted
independently from the boundary scalar equation
of state.

The paper is organized as follows.
Sec.~\ref{sec:model} introduces the models and
the corresponding black brane solutions.
Sec.~\ref{sec:firstlaw} establishes the
thermodynamic relation between the boundary
and the black hole interior.
Sec.~\ref{sec:cancellation} examines the
behavior of the interior geometry near
the onset of scalarization.
Sec.~\ref{sec:landau} develops the interior
Landau theory and compares its predictions
with the numerical solutions and the
boundary thermodynamics.
We conclude in Sec.~\ref{sec:discussion}.

\section{Model and black brane solutions}
\label{sec:model}

\subsection{Model and thermodynamic ensemble}
\label{subsec:setup}

We consider the Einstein--Maxwell theory coupled to a real scalar,
\begin{equation}
 I=\frac{1}{16\pi G_N}\int\dd^4x\sqrt{-g}\left[
 R-\frac14G(\phi)F_{ab}F^{ab}
 -\frac12(\nabla\phi)^2-V(\phi)\right],
 \qquad F=\dd A.
 \label{eq:action}
\end{equation}
We set the AdS radius $L=1$ and $16\pi G_N=1$.  The two models
of Ref.~\cite{GaoLiuLyu2609} used below are
\begin{equation}
 \begin{aligned}
 \calM_{\rm I}:\quad
 &G(\phi)=1+0.6\phi^2,
 &V(\phi)&=-6-\phi^2+10\phi^4,\\
 \calM_{\rm II}:\quad
 &G(\phi)=1+0.6\phi^2,
 &V(\phi)&=-6-\phi^2.
 \end{aligned}
 \label{eq:models}
\end{equation}
Both models have $m^2=-2$ and identical linear scalar equations
around the RN background. They therefore share the same critical
temperature but differ in their nonlinear scalar interactions.
Comparing them allows us to examine the effects of these interactions
on the phase transition and the terminal Kasner geometry.

For homogeneous solutions we take the ansatz
\begin{equation}
 \begin{split}
 \dd s^2&=\frac{1}{z^2}\left[-f(z)e^{-\chi(z)}\dd t^2
 +\frac{\dd z^2}{f(z)}+\dd x^2+\dd y^2\right],\\
 A&=\psi(z)\dd t,\qquad \phi=\phi(z).
 \end{split}
 \label{eq:ansatz}
\end{equation}
The AdS boundary lies at $z=0$, with increasing $z$ pointing
into the bulk.  We set $\chi(0)=0$ to fix the boundary time
normalization.  The outer event horizon is at $z=z_h$, where
$f(z_h)=0$, and we choose the regular gauge $\psi(z_h)=0$.
The Maxwell equation has the first integral
\begin{equation}
 G(\phi)e^{\chi/2}\psi'=-\rho,
 \label{eq:maxwell}
\end{equation}
where $\rho$ is the conserved boundary charge density. Here and
throughout, a prime denotes differentiation with respect to $z$,
while a subscript $,\phi$ denotes differentiation with respect to
$\phi$. The remaining equations of motion are
\begin{align}
 \chi'&=\frac z2(\phi')^2,
 \label{eq:chi_eq}\\
 \left(\frac{e^{-\chi/2}f}{z^3}\right)'
 &=\frac{e^{-\chi/2}}{4z^4}
 \left[\frac{z^4\rho^2}{G(\phi)}+2V(\phi)\right],
 \label{eq:f_eq}\\
 \left(\frac{e^{-\chi/2}f\phi'}{z^2}\right)'
 &=\frac{e^{-\chi/2}}{z^4}\left[
 V_{,\phi}-\frac{z^4\rho^2G_{,\phi}}{2G(\phi)^2}\right].
 \label{eq:scalar_eq}
\end{align}
The Hawking temperature and entropy density are
\begin{equation}
 T=-\frac{e^{-\chi_h/2}f'(z_h)}{4\pi},
 \qquad s=\frac{4\pi}{z_h^2}.
 \label{eq:Ts}
\end{equation}

Near the AdS boundary, the scalar has the expansion
\begin{equation}
 \phi(z)=Jz+Oz^2+\cdots.
 \label{eq:uvphi}
\end{equation}
We use standard quantization, in which $J$ is the source
and $O$ the expectation value of the dual scalar operator.
The other fields obey
\begin{equation}
 \psi(z)=\mu-\rho z+\cdots,\qquad
 f(z)=1+\frac{J^2}{4}z^2+f_3z^3+\cdots.
 \label{eq:uvmetric}
\end{equation}
We work in the grand-canonical ensemble at fixed chemical
potential $\mu$ and vanishing scalar source $J=0$.
By varying the temperature, we compare the grand potentials
of the RN and scalarized phases to determine the
thermodynamically preferred phase.
Near the critical temperature, we also consider regular
solutions with $J\ne0$ to reconstruct the local equation of state.

Holographic renormalization in standard quantization gives the grand potential and corresponding first law
\cite{GaoLiuLyu2609,Skenderis2002}
\begin{align}
    \calF &= f_3-JO,
    \label{eq:freeUV}\\
    \dd\calF &= -s\,\dd T-\rho\,\dd\mu-O\,\dd J.
    \label{eq:dF}
\end{align}
Since $\calF$ has scaling dimension three, while
$T$, $\mu$, and $J$ have dimension one, the first law
implies the Smarr relation
\begin{equation}
    3f_3-2JO+\mu\rho+Ts=0.
    \label{eq:ward}
\end{equation}

\subsection{RN and scalarized black branes}
\label{subsec:terminal}

The normal phase is described by the planar
RN--AdS black brane,
\begin{equation}
    \begin{split}
        f_{\RN}(z)
        &=1-\left(\frac{1}{z_h^3}
        +\frac{\mu^2}{4z_h}\right)z^3
        +\frac{\mu^2}{4z_h^2}z^4,\\
        \psi_{\RN}(z)
        &=\mu\left(1-\frac{z}{z_h}\right),\\
        \phi_{\RN}(z)&=0,\qquad \chi_{\RN}(z)=0.
    \end{split}
    \label{eq:RN}
\end{equation}
The Hawking temperature and grand potential of the RN black brane
are given by
\begin{equation}
    T_{\RN}=\frac{3-\mu^2z_h^2/4}{4\pi z_h},
    \qquad
    \calF_{\RN}=-\frac{1}{z_h^3}-\frac{\mu^2}{4z_h}.
    \label{eq:RNthermo}
\end{equation}
At given $(T,\mu)$, the outer-horizon radius $z_+$ is
determined by the temperature relation. For a nonextremal
RN black brane, $f_{\RN}(z)$ has another zero at $z=z_->z_h$,
corresponding to the inner Cauchy horizon. 

The scalarized solutions obey the same AdS boundary conditions,
but their interior geometry is different. Scalar backreaction
destroys the RN Cauchy horizon, and the solutions
approach a scalar-kinetic-dominated Kasner singularity as
$z\to\infty$ \cite{GaoLiuLyu2609}. Near the singularity, the fields behave as
\begin{align}
 \phi(z)&=v\log z+\phi_{\K}+o(1),
 \label{eq:Kasner_phi}\\
 \chi(z)&=\frac{v^2}{2}\log z+\chi_{\K}+o(1),
 \label{eq:Kasner_chi}\\
 f(z)&=-f_{\K}z^{3+v^2/4}\bigl(1+o(1)\bigr),
 \label{eq:Kasner_f}\\
 \psi(z)&=\psi_{\K}+o(1).
 \label{eq:Kasner_psi}
\end{align}
Here $v$ describes the logarithmic growth of the scalar. Introducing the proper time $\tau\propto z^{-(v^2+12)/8}$
near the singularity, the metric and scalar field take
the Kasner form
\begin{equation}
    \begin{split}
    \dd s^2 &\simeq -\dd\tau^2
    +\tau^{2p_t}\dd t^2
    +\tau^{2p_x}\dd x^2
    +\tau^{2p_y}\dd y^2,\\
    \phi &\simeq \sqrt{2}\,p_\phi\log\tau+O(1).
    \end{split}
    \label{eq:Kasner_proper}
\end{equation}
The corresponding Kasner exponents are
\begin{equation}
    p_t=\frac{v^2-4}{v^2+12},
    \qquad
    p_x=p_y=\frac{8}{v^2+12},
    \qquad
    p_\phi=-\frac{4\sqrt{2}v}{v^2+12}.
    \label{eq:kasner_exponents}
\end{equation}
They satisfy the generalized Kasner constraints
$p_t+p_x+p_y=1$ and
$p_t^2+p_x^2+p_y^2+ p_\phi^2=1$.

For the polynomial couplings in \eqref{eq:models},
a Kasner-like plateau is not necessarily the final
asymptotic regime: electric backreaction can induce a later
Kasner inversion.  In particular, the finite limit
$\psi_{\K}$ assumed in \eqref{eq:Kasner_psi} is
consistent with the terminal scalar-kinetic regime when
$|v|>2$.  All endpoint formulas below refer to the
terminal regime, not an intermediate plateau.

\section{Thermodynamics from the interior}
\label{sec:firstlaw}

The RN inner horizon and the terminal Kasner singularity
provide two distinct interior surfaces for evaluating
the thermodynamics of the corresponding black branes.
In the grand-canonical ensemble, their relative stability
is determined by the grand potential at fixed $(T,\mu)$.
A radial conservation law connects the outer-horizon
thermodynamics to either interior endpoint.

For the ansatz \eqref{eq:ansatz}, the Einstein and Maxwell
equations combine into
\begin{equation}
 \left[\frac{e^{-\chi/2}}{z^2}(f'-f\chi')\right]'
 =\frac{\rho^2e^{-\chi/2}}{G(\phi)}
 =-\rho\psi'.
 \label{eq:scaling_derivative}
\end{equation}
Integrating this relation gives a conserved charge
\begin{equation}
 \calQ=\frac{e^{-\chi/2}}{z^2}(f'-f\chi')+\rho\psi,
 \qquad \calQ'=0.
 \label{eq:Q}
\end{equation}
At the outer horizon, where $\psi(z_h)=0$, this charge
equals $\calQ_h=-Ts$.  Because $\mu$ and the electric flux
$\rho$ are constant along each solution, it is convenient
to use the shifted invariant
\begin{equation}
 \calK\equiv\mu\rho-\calQ=Ts+\mu\rho.
 \label{eq:Kthermo}
\end{equation}
We can therefore evaluate $Ts+\mu\rho$ at either
interior endpoint.

For a scalarized black brane, introduce the finite metric
amplitude
\begin{equation}
    \Xi\equiv-\lim_{z\to\infty}\frac{e^{-\chi/2}f}{z^3}
    =f_{\K}e^{-\chi_{\K}/2}.
    \label{eq:Xi}
\end{equation}
Substituting the terminal Kasner asymptotics
\eqref{eq:Kasner_phi}--\eqref{eq:Kasner_psi}
into \eqref{eq:Q} and~\eqref{eq:Kthermo} gives
\begin{equation}
    \calK_{\rm H}
    =\Xi\left(3-\frac{v^2}{4}\right)
    +\rho_{\rm H}(\mu-\psi_{\K}).
    \label{eq:Kcomponents}
\end{equation}
The first and second terms are respectively the metric
and electric contributions to the same conserved quantity.

In the normal phase the corresponding endpoint is the RN
inner Cauchy horizon $z_->z_h$.  We define its signed
temperature, entropy density and electric potential by
\begin{equation}
 T_-=-\frac{f'_{\RN}(z_-)}{4\pi},
 \qquad s_-=\frac{4\pi}{z_-^2},
 \qquad \psi_-=\mu-\rho_{\RN}z_-.
 \label{eq:rn_inner_Ts}
\end{equation}
Note that $T_-<0$ for the nonextremal RN solution.
Evaluating Eq.~\eqref{eq:Q} at $z_-$ gives
\begin{equation}
    \calK_{\RN}
    =T_-s_-+\rho_{\RN}(\mu-\psi_-).
    \label{eq:RN_inner_K}
\end{equation}
Eqs.~\eqref{eq:Kcomponents}
and~\eqref{eq:RN_inner_K} are the two endpoint
evaluations of the same radial invariant, on different
black brane solutions.

The boundary Smarr identity \eqref{eq:ward} and
$\calF=f_3-JO$ imply
\begin{equation}
 \calF(T,\mu,J)=-\frac{\calK+JO}{3}.
 \label{eq:Fgeneral}
\end{equation}
For the source-free phases, the grand potentials can
therefore be read directly from their interior endpoints:
\begin{equation}
 \begin{split}
 \calF_{\RN}(T,\mu)
 &=-\frac13\left[T_-s_-
 +\rho_{\RN}(\mu-\psi_-)\right],\\
 \calF_{\rm H}(T,\mu)
 &=-\frac13\left[\Xi\left(3-\frac{v^2}{4}\right)
 +\rho_{\rm H}(\mu-\psi_{\K})\right].
 \end{split}
 \label{eq:grand_compare}
\end{equation}
Both expressions are evaluated at the same $(T,\mu)$,
with the respective charges and geometries determined
on each branch.  The smaller grand potential identifies
the thermodynamically preferred phase. 

\begin{figure}[h!]
 \centering
 \includegraphics[width=0.9\textwidth]{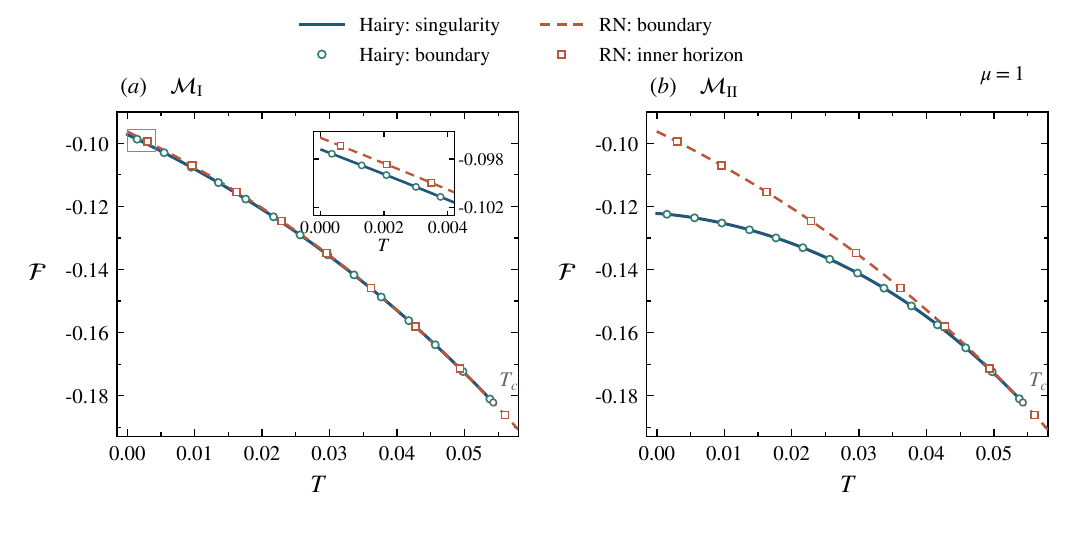}
 \caption{\small Grand potentials of the RN--AdS and scalarized black branes
for (a) model I and (b) model II at $\mu=1$.
For the scalarized solutions, the solid curves are obtained
from the terminal Kasner singularity, while the circles
represent independent evaluations at the AdS boundary.
For RN--AdS, the dashed curves and squares correspond to
the boundary and inner-horizon evaluations, respectively. The inset in panel (a) resolves the small
free-energy separation at low temperatures.}
 \label{fig:free_energy_comparison}
\end{figure}

Fig.~\ref{fig:free_energy_comparison} shows the grand
potentials of both phases for the two models.
The evaluations at the RN inner horizon and the terminal
Kasner singularity agree with the corresponding boundary
results. In both models, the scalarized branch lies below
the RN--AdS branch for $T<T_c$, establishing its
thermodynamic preference directly from the interior
expressions \eqref{eq:grand_compare}.

\section{Critical matching and quadratic response}
\label{sec:cancellation}

The RN and scalarized black branes have different interior
endpoints: a Cauchy horizon in the normal phase and a terminal
Kasner singularity in the scalarized phase. As the scalarized
branch approaches the continuous transition, its exterior
geometry converges to the critical RN solution. The radial
invariant $\calK=Ts+\mu\rho$ therefore has the same limit
on both branches,
\begin{equation}
 \lim_{T\to T_c}\calK_{\rm H}
 =\calK_{\RN,c}=T_cs_c+\mu\rho_c,
 \label{eq:critical_K_limit}
\end{equation}
where the limit is taken along the source-free scalarized
branch at fixed $\mu$.

\begin{figure}[h!]
 \centering
 \includegraphics[width=0.9\textwidth]
 {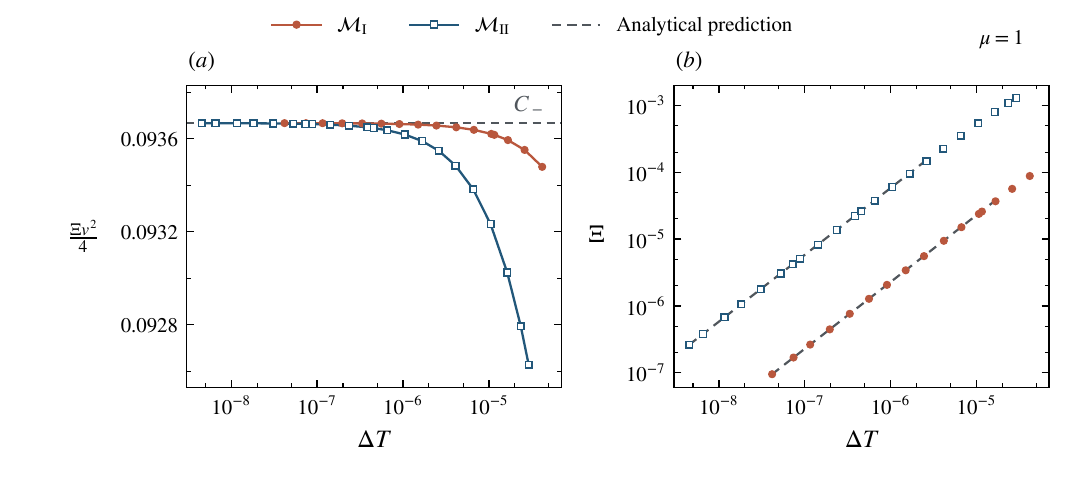}
 \caption{\small Near-critical behavior of the terminal Kasner geometry
along the source-free scalarized branches at $\mu=1$,
with $\Delta T=T_c-T$.
(a) The combination $\Xi v^2/4$ approaches the
critical RN inner-horizon value $C_-=-T_-s_-$
(horizontal dashed line).
(b) The Kasner metric amplitude $\Xi$ vanishes
linearly as $\Delta T\to0$, in agreement with
the analytical predictions (dashed lines).}
 \label{fig:critical_limit}
\end{figure}

The approach to this limit is nonuniform in the interior.
In both models, the terminal Kasner solutions exhibit
\begin{equation}
 \lim_{T\to T_c}\Xi=0,
 \qquad
 \lim_{T\to T_c}\frac{\Xi v^2}{4}
 =C_-,\qquad
 C_-\equiv\frac{f'_{\RN}(z_-)}{z_-^2}=-T_-s_-,
 \label{eq:two_phase_finite_limit}
\end{equation}
where the RN quantities on the right are evaluated at
$(T_c,\mu)$. As a result, the metric term in the Kasner
expression \eqref{eq:Kcomponents} approaches $T_-s_-$,
the corresponding term at the RN inner horizon. Together
with Eq.~\eqref{eq:critical_K_limit}, this also fixes
the limiting electric contribution to the value in
Eq.~\eqref{eq:RN_inner_K}. The numerical approach to
these limits is shown in Fig.~\ref{fig:critical_limit}.

The limiting value of $\calK$ does not determine its
quadratic response to scalar hair. We examine this
response by perturbing the critical RN solution at
$T=T_c$ and fixed $\mu$, using the horizon scalar
amplitude $\eps=\phi_h$ as the perturbation parameter.

Let $z_c$ be the outer-horizon radius of the critical
RN black brane, with $\rho_c=\mu/z_c$, and write
$f_{\RN,c}(z)$ for its metric function.
The leading scalar perturbation is
$\phi(z)=\eps\Phi(z)+\mathcal O(\eps^3)$.
At criticality, $\Phi$ is a normalizable zero mode,
$\Phi(z_c)=1$ and $\Phi(z)\sim z^2$ as $z\to0$,
which satisfies
\begin{equation}
 \left(\frac{f_{\RN,c}}{z^2}\Phi'\right)'
 +\left(\frac{2}{z^4}+g_2\rho_c^2\right)\Phi=0,
 \qquad g_2=0.6.
 \label{eq:zeroMode}
\end{equation}
Here and below primes denote derivatives with respect
to $z$. The equation depends only on the quadratic
scalar coupling and is the same in both models.
Since $\Phi$ is normalizable, the linear scalar
perturbation has vanishing boundary source.
By the $\phi\to-\phi$ symmetry, a nonzero source
can first appear at $\mathcal O(\eps^3)$ and therefore
does not affect the quadratic response considered below.

The scalar backreaction
changes the entropy and charge density
at even orders in $\eps$,
\begin{equation}
 s=s_c+s_2\eps^2+\mathcal O(\eps^4),\qquad
 \rho=\rho_c+\rho_2\eps^2+\mathcal O(\eps^4).
 \label{eq:critical_sr}
\end{equation}
Eq.~\eqref{eq:Kthermo} then gives
\begin{equation}
 \calK(T_c,\mu,\eps)
 =\calK_{\RN,c}+(T_cs_2+\mu\rho_2)\eps^2
 +\mathcal O(\eps^4).
 \label{eq:K_sourced_second_order}
\end{equation}
The coefficient $T_cs_2+\mu\rho_2$ can be determined
from the Maxwell equation and the horizon temperature,
without solving for the full second-order metric.
We expand the horizon position and redshift function as
\begin{equation}
 z_h=z_c+z_2\eps^2+\mathcal O(\eps^4),\qquad
 \chi(z)=\eps^2\chi_2(z)+\mathcal O(\eps^4),\qquad
 \chi_2'(z)=\frac z2(\Phi')^2,
 \label{eq:backreac}
\end{equation}
with $\chi_2(0)=0$.
Integrating the Maxwell equation~\eqref{eq:maxwell}
from the boundary to the horizon, where $\psi(z_h)=0$,
gives
\begin{equation}
 \mu=\rho\int_0^{z_h}\dd z\,\frac{e^{-\chi/2}}{G(\phi)}
 =\rho\left[z_c+(z_2-I_\Phi)\eps^2\right]
 +\mathcal O(\eps^4),
 \label{eq:maxwell_expansion}
\end{equation}
where $G(\phi)=1+g_2\phi^2+\mathcal O(\phi^4)$ and
\begin{equation}
 I_\Phi=\int_0^{z_c}\dd z\left[
 \frac{\chi_2(z)}{2}+g_2\Phi^2\right]
 =\int_0^{z_c}\dd z\left[
 \frac{z(z_c-z)}{4}(\Phi')^2+g_2\Phi^2\right].
 \label{eq:Iphi}
\end{equation}
Since $\mu$ is fixed and $\rho_c=\mu/z_c$,
Eq.~\eqref{eq:maxwell_expansion} then gives
\begin{equation}
 \rho_2=\frac{\mu}{z_c^2}(I_\Phi-z_2).
 \label{eq:rho2}
\end{equation}

The horizon displacement follows from the fixed-temperature
condition. Evaluating \eqref{eq:f_eq} at the
outer horizon $f(z_h)=0$ gives
\begin{equation}
 4\pi T=-e^{-\chi(z_h)/2}
 \left[\frac{\rho^2z_h^3}{4G(\phi_h)}
       +\frac{V(\phi_h)}{2z_h}\right].
 \label{eq:critical_horizon_temp}
\end{equation}
Here $\phi_h=\eps$, so that
$G(\phi_h)=1+g_2\eps^2+\mathcal O(\eps^4)$ and
$V(\phi_h)=-6-\eps^2+\mathcal O(\eps^4)$.
Expanding~\eqref{eq:critical_horizon_temp}
at fixed $T=T_c$ yields a relation between the
horizon displacement and charge correction,
\begin{equation}
 \left(\frac{3}{z_c^2}+\frac{3\mu^2}{4}\right)z_2
 +\frac{\mu z_c^2}{2}\rho_2
 =\frac{\mu^2g_2z_c}{4}+\frac{1}{2z_c}
 -2\pi T_c\chi_2(z_c).
 \label{eq:temperature_shift}
\end{equation}
The right-hand side can be expressed in terms of
$I_\Phi$ using the critical zero-mode equation.
Introduce the combination
\begin{equation}
 \mathcal J_\Phi=-\frac{f_{\RN,c}}{z}(\Phi')^2
 -\left(\frac{2}{z^3}+g_2\rho_c^2z\right)\Phi^2
 +\frac{3f_{\RN,c}}{z^2}\Phi\Phi'.
 \label{eq:critical_zero_mode_current}
\end{equation}
Eq.~\eqref{eq:zeroMode} implies
\begin{equation}
 \mathcal J_\Phi'=\frac{f_{\RN,c}'}{z}(\Phi')^2
 -4g_2\rho_c^2\Phi^2.
 \label{eq:critical_current_derivative}
\end{equation}
The zero-mode boundary conditions give
$\mathcal J_\Phi(0)=0$ and
$\mathcal J_\Phi(z_c)=-2/z_c^3-g_2\rho_c^2z_c$.
Integrating \eqref{eq:critical_current_derivative}
and using
$-f_{\RN,c}'/z=\rho_c^2z(z_c-z)+4\pi T_cz/z_c^2$
together with \eqref{eq:backreac} and
\eqref{eq:Iphi}, we find
\begin{equation}
 \mu^2I_\Phi
 =\frac{\mu^2g_2z_c}{4}
 +\frac{1}{2z_c}-2\pi T_c\chi_2(z_c).
 \label{eq:virialphysical}
\end{equation}
Combining this identity with \eqref{eq:rho2}
and \eqref{eq:temperature_shift} gives
\begin{equation}
 \left(\frac{3}{z_c^2}+\frac{\mu^2}{4}\right)z_2
 =\frac{\mu^2}{2}I_\Phi.
 \label{eq:temperature_shift_reduced}
\end{equation}

Finally, $s=4\pi/z_h^2$ gives
$s_2=-8\pi z_2/z_c^3$. Combining this with
\eqref{eq:rho2} and the critical RN relation
$4\pi T_c=3/z_c-\mu^2z_c/4$, we obtain
\begin{equation}
 T_cs_2+\mu\rho_2
 =\frac{\mu^2I_\Phi}{z_c^2}
 -\frac{12+\mu^2z_c^2}{2z_c^4}z_2=0,
 \label{eq:cancel}
\end{equation}
where the last equality follows from
\eqref{eq:temperature_shift_reduced}. Hence
\begin{equation}
 \calK(T_c,\mu,\eps)
 =\calK_{\RN,c}+\mathcal O(\eps^4).
 \label{eq:K4eps}
\end{equation}
Since the calculation involves only the common
linearized scalar equation and its second-order
backreaction, the cancellation holds in both models.

\section{Landau theory from the Kasner endpoint}
\label{sec:landau}

The preceding sections establish that the radial
invariant $\calK$ determines the grand potential
of the source-free phases and has a vanishing
quadratic response at the critical point.
We now construct a Landau potential from its
response along regular solutions with nonzero
scalar source at fixed $(T,\mu)$.
The resulting theory will then be tested on the
physical $J=0$ branches and compared with the
conventional boundary Landau description.

\subsection{Interior Landau potential}
\label{subsec:interior_landau_potential}

\begin{figure}[t]
 \centering
 \includegraphics[width=0.9\textwidth]
 {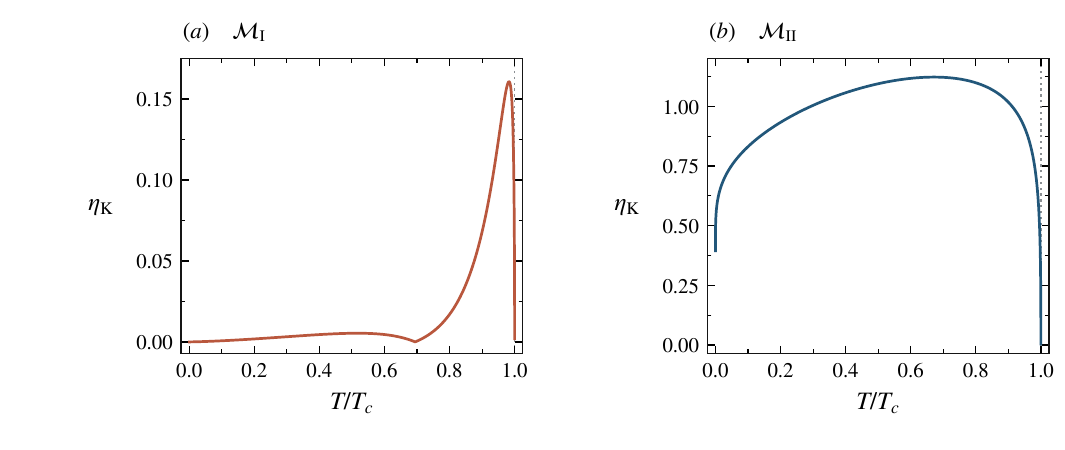}
 \caption{\small
 Temperature dependence of $\eta_{\K}=4/v$
 along the source-free scalarized branches at
 $\mu=1$ for (a) model I and (b) model II.
 The dotted vertical lines mark $T=T_c$.
 }
 \label{fig:kasner_order_temperature}
\end{figure}

We introduce the signed interior order parameter
\begin{equation}
 \eta_{\K}=\frac{4}{v},
 \label{eq:eta}
\end{equation}
which vanishes at the RN bifurcation.
Under $\phi\to-\phi$, the Kasner velocity $v$
changes sign, whereas $\calK$ remains invariant.
Figure~\ref{fig:kasner_order_temperature} shows
$\eta_{\K}$ along the source-free scalarized branches.
It approaches zero at $T_c$ in both models, but its
temperature dependence away from the critical point
is different.

For the regular sourced solutions at fixed $(T,\mu)$,
we assume a smooth, nondegenerate expansion near
the bifurcation. The endpoint response is even in
$\eta_{\K}$ and takes the form
\begin{equation}
 \calK(T,\mu,\eta_{\K})
 =\calK_{\RN}(T,\mu)
 +a_{\K}(T,\mu)\eta_{\K}^{2}
 +b_{\K}(T,\mu)\eta_{\K}^{4}
 +\mathcal O(\eta_{\K}^{6}).
 \label{eq:KLandauExpansion}
\end{equation}
Here $\calK_{\RN}$ denotes the normal-phase
value at the same thermodynamic parameters.

At fixed $(T_c,\mu)$, the scalar source enters
at $\mathcal O(\eps^3)$ and does not affect the
leading critical zero mode.
The Cauchy-horizon matching of
Ref.~\cite{GaoLiuLyu2609} therefore gives
$\eta_{\K}\propto\eps$ for the sourced family.
Together with the quadratic cancellation
in \eqref{eq:K4eps}, this implies
$a_{\K}(T_c,\mu)=0$.
We define
\begin{equation}
 \alpha_{\K}
 =\left.\frac{\partial a_{\K}}{\partial T}
 \right|_{T_c,\mu},
 \qquad
 b_{\K}^{c}=b_{\K}(T_c,\mu).
 \label{eq:alphaBetaK}
\end{equation}

To obtain the corresponding potential,
we use the Iyer--Wald first law derived in
Appendix~\ref{app:iyerwald}.
For variations between regular sourced
solutions at fixed $(T,\mu)$, it gives
\begin{equation}
 \begin{split}
 \dd\calK
 &=
 \Xi\left(2\,\dd\log f_{\K}
 -v\,\dd\phi_{\K}\right)
 +(\mu-\psi_{\K})\,\dd\rho\\
 &=2O\,\dd J-J\,\dd O.
 \end{split}
 \label{eq:masterfixed}
\end{equation}
Thus the response of the terminal Kasner
geometry is directly related to the
boundary scalar source.

In standard quantization, the Legendre-transformed
grand potential
$\Gamma(T,\mu,O)=\calF(T,\mu,J)+JO$
satisfies $\partial_O\Gamma=J$ at fixed
$(T,\mu)$.
Together with Eq.~\eqref{eq:Fgeneral},
this yields
\begin{equation}
 2O\left(\frac{\partial\Gamma}{\partial O}
 \right)_{T,\mu}
 -3\Gamma=\calK.
 \label{eq:LandauODE}
\end{equation}

Since the boundary condensate is also linear
in $\eps$ near the bifurcation, we have
\begin{equation}
 O=c\eta_{\K}
 +\mathcal O\!\left(
 \eta_{\K}^3,(T-T_c)\eta_{\K}\right),
 \qquad c\ne0.
 \label{eq:boundary_interior_map}
\end{equation}
Defining
$\calV_{\K}(T,\mu,\eta_{\K})
=\Gamma(T,\mu,O(\eta_{\K}))$
and using
$O/(\partial_{\eta_{\K}}O)
=\eta_{\K}+\mathcal O(\eta_{\K}^3)$,
Eq.~\eqref{eq:LandauODE} gives
\begin{equation}
 \begin{split}
 \calV_{\K}(T,\mu,\eta_{\K})
 ={}&\calF_{\RN}(T,\mu)
 +\alpha_{\K}(T-T_c)\eta_{\K}^{2}\\
 &+\frac{b_{\K}^{c}}{5}\eta_{\K}^{4}
 +\mathcal O\!\left(
 (T-T_c)^2\eta_{\K}^{2},
 (T-T_c)\eta_{\K}^{4},
 \eta_{\K}^{6}\right).
 \end{split}
 \label{eq:interior_landau_normal_form}
\end{equation}
This gives the interior Landau potential constructed
from the terminal Kasner response.
Its stationary points coincide with the physical
source-free branches, since
$\partial_{\eta_{\K}}\calV_{\K}
=J\,\partial_{\eta_{\K}}O$.

\subsection{Critical behavior and numerical verification}
\label{subsec:interior_landau_tests}

We extract the coefficients of the interior
potential from regular sourced solutions.
At fixed $(T,\mu)$, the horizon scalar amplitude
is varied, and the resulting Kasner values of
$\calK$ and $\eta_{\K}$ are fitted to
Eq.~\eqref{eq:KLandauExpansion}.
For $\mu=1$, the zero of $a_{\K}(T)$ occurs at
$T\simeq0.0542734$ in both models, consistent
with $T_c\simeq0.05427335$ obtained from the
normalizable scalar zero mode of RN \cite{GaoLiuLyu2609}.
The fitted coefficients are given in
Table~\ref{tab:interior_coefficients}.

\begin{table}[t]
 \centering
 \begin{tabular}{lcccc}
 \toprule
 Model & $\alpha_{\K}$ & $b_{\K}^{c}$
 & $5\alpha_{\K}/(2b_{\K}^{c})$
 & $5\alpha_{\K}^{2}/(4b_{\K}^{c})$\\
 \midrule
 $\calM_{\rm I}$
 & $8.64\times10^{-3}$
 & $2.23\times10^{-4}$
 & $97.1$ & $0.419$\\
 $\calM_{\rm II}$
 & $8.64\times10^{-3}$
 & $8.84\times10^{-6}$
 & $2.44\times10^3$ & $10.56$\\
 \bottomrule
 \end{tabular}
 \caption{\small
 Interior Landau coefficients from the sourced
 Kasner endpoints at $\mu=1$, together with
 the amplitudes entering
 Eq.~\eqref{eq:interior_landau_predictions}.
 The quartic coefficients have a few-percent
 sensitivity to the extrapolation procedure.
 }
 \label{tab:interior_coefficients}
\end{table}

\begin{figure}[t]
 \centering
 \includegraphics[width=0.9\textwidth]
 {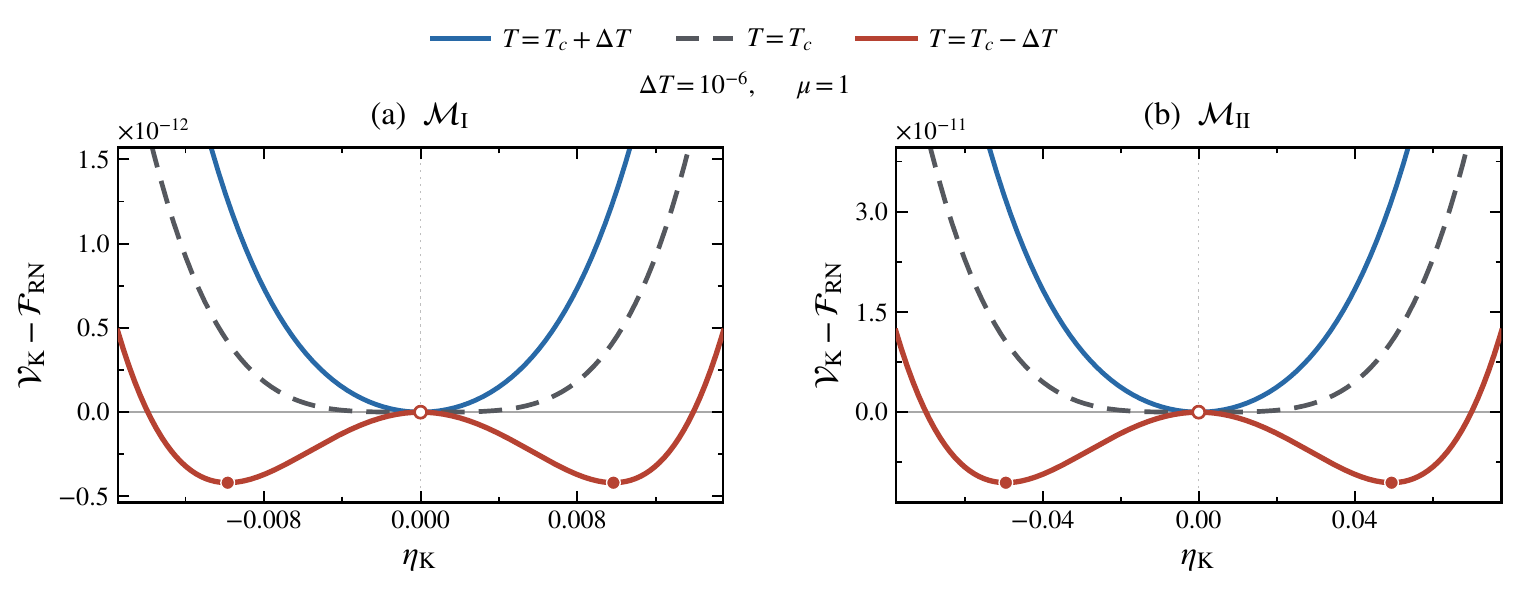}
 \caption{\small
 Interior Landau potentials for (a) model I and
 (b) model II at $\mu=1$, evaluated at
 $T=T_c+\Delta T$, $T_c$, and $T_c-\Delta T$
 with $\Delta T=10^{-6}$. The filled circles
 mark the nonzero minima below $T_c$.
 }
 \label{fig:landau_potential}
\end{figure}

The coefficients $\alpha_{\K}$ and $b_{\K}^{c}$
are positive in both models. Hence the potential
\eqref{eq:interior_landau_normal_form} has its
minimum at $\eta_{\K}=0$ for $T>T_c$, whereas
two symmetry-related minima appear for $T<T_c$.
Minimization gives
\begin{equation}
 \begin{split}
 \eta_{\K,\mathrm{eq}}^{2}
 &=\frac{5\alpha_{\K}}{2b_{\K}^{c}}
 (T_c-T)+\mathcal O((T_c-T)^2),\\
 \Delta\calF
 &\equiv\calF_{\RN}-\calF_{\rm H}
 =\frac{5\alpha_{\K}^{2}}{4b_{\K}^{c}}
 (T_c-T)^2+o((T_c-T)^2).
 \end{split}
 \label{eq:interior_landau_predictions}
\end{equation}
The first relation gives the mean-field exponent
$\beta=1/2$ for the boundary condensate through
\eqref{eq:boundary_interior_map}; the second
fixes the leading grand-potential difference.
Fig.~\ref{fig:landau_potential} shows the resulting
potentials. The marked difference between the
two models comes from their quartic coefficients.

\begin{figure}[t]
 \centering
 \includegraphics[width=0.9\textwidth]
 {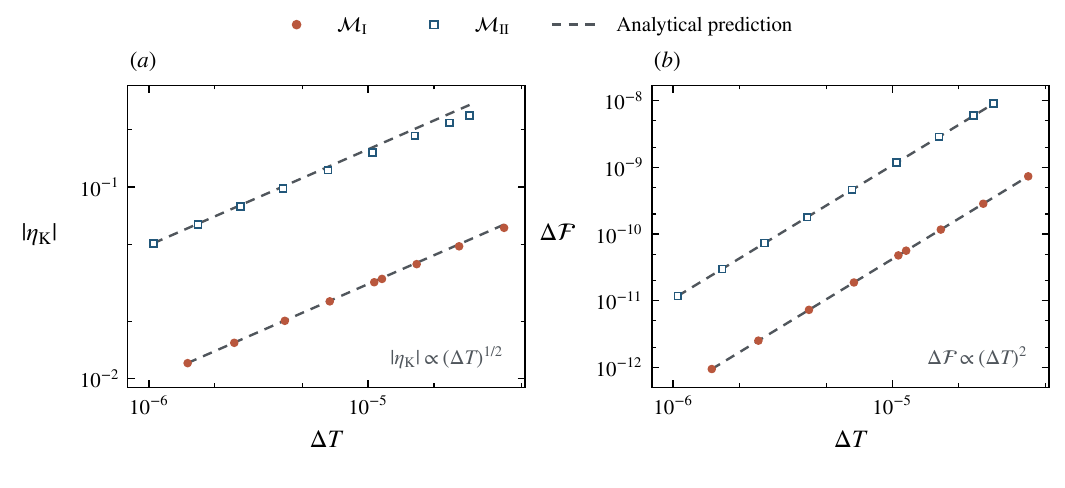}
 \caption{\small
 Critical behavior of the source-free scalarized
 branches at $\mu=1$: (a) the order parameter
 $|\eta_{\K}|$ and (b) the grand-potential
 difference $\Delta\calF=\calF_{\RN}-\calF_{\rm H}$,
 against $\Delta T=T_c-T$. Symbols denote numerical
 solutions and the curves follow
 \eqref{eq:interior_landau_predictions},
 with coefficients obtained from the sourced
 Kasner endpoints.
 }
 \label{fig:interior_validation}
\end{figure}

We test these predictions on the physical
source-free branches, using $\eta_{\K}$ from
the terminal Kasner geometry and the
grand-potential difference from
\eqref{eq:grand_compare}.
The numerical results in
Fig.~\ref{fig:interior_validation} agree with
\eqref{eq:interior_landau_predictions},
including the amplitudes fixed by the sourced
endpoint coefficients.

\subsection{Interior--boundary coefficient matching}
\label{subsec:landau_correspondence}

The boundary scalar equation of state provides
an independent test of the Landau potential
constructed from the Kasner endpoint.
The $\phi\to-\phi$ symmetry makes $J$ odd in $O$,
while the normalizable zero mode at $T_c$
requires its linear coefficient to vanish
at criticality. Near the bifurcation, we have
\begin{equation}
 J=A(T-T_c)O+B_{\calM}O^3
 +\mathcal O\!\left(
 (T-T_c)^2O,(T-T_c)O^3,O^5\right).
 \label{eq:boundary_landau_EOS}
\end{equation}
Integrating $\partial_O\Gamma=J$
at fixed $(T,\mu)$ gives
\begin{equation}
 \Gamma(T,\mu,O)
 =\calF_{\RN}(T,\mu)
 +\frac{A}{2}(T-T_c)O^2
 +\frac{B_{\calM}}{4}O^4+\cdots.
 \label{eq:boundary_landau_potential}
\end{equation}

Substituting \eqref{eq:boundary_interior_map}
into the boundary potential
\eqref{eq:boundary_landau_potential}
and comparing with the interior potential
\eqref{eq:interior_landau_normal_form},
we obtain
\begin{equation}
 \alpha_{\K}=\frac{Ac^2}{2},
 \qquad
 b_{\K}^{c}=\frac{5B_{\calM}c^4}{4}.
 \label{eq:landau_coefficient_matching}
\end{equation}

At $\mu=1$, fitting the boundary equation
of state on the regular sourced solutions
gives $A\simeq60.335$ in both models,
$B_{\calM_{\rm I}}\simeq2163.8$ and
$B_{\calM_{\rm II}}\simeq84.896$.
The corresponding coefficients relating
the two order parameters are
$c_{\calM_{\rm I}}\simeq0.016923$ and
$c_{\calM_{\rm II}}\simeq0.016924$.
Eq.~\eqref{eq:landau_coefficient_matching}
then gives $\alpha_{\K}\simeq8.6396\times10^{-3}$
and $8.6402\times10^{-3}$, and
$b_{\K}^{c}\simeq2.2184\times10^{-4}$
and $8.7050\times10^{-6}$,
for models I and II, respectively.
These values agree well with the Kasner endpoint
coefficients in Table~\ref{tab:interior_coefficients}.

\section{Conclusions and discussion}
\label{sec:discussion}

We have constructed a Landau theory of a continuous
black-brane phase transition from the terminal
Kasner geometry. The conserved radial charge
determines the grand potential from either the
RN inner horizon or the Kasner singularity.
At the critical point, the quadratic response
of this charge vanishes analytically.
Together with the Iyer--Wald first law,
this allows the leading Landau potential
to be determined entirely from the
Kasner endpoint, without using the
boundary scalar equation of state.

The two models considered here share the same
critical temperature and leading thermal
Landau coefficient, but have substantially
different quartic responses due to their
nonlinear scalar interactions.
The resulting predictions for the Kasner
order parameter and the grand-potential
difference agree with the fully backreacted
black-brane solutions.
The independently extracted boundary
Landau coefficients also agree with the
interior results.
These findings demonstrate that the terminal
Kasner geometry encodes not only the critical
scaling, but also the model-dependent
thermodynamics of the transition.

The present construction is restricted to homogeneous
black branes near a continuous phase transition.
Extending it to spatially varying configurations
would require incorporating gradient terms into the
interior Landau potential, thereby establishing a
Ginzburg--Landau description that may also capture
spatial correlations and domain-wall configurations.
Beyond the critical region, the nonlinear relation
between the boundary condensate and the Kasner order
parameter becomes important, particularly for
first-order and multicritical transitions.
It remains to be understood how this construction
extends to other scalar boundary conditions and
thermodynamic ensembles, or to interiors with
multiple Kasner epochs. A further question is
whether the interior Landau description can be
connected to dynamical critical behavior, including
the soft quasinormal modes near the transition.

\acknowledgments

We thank Hong L\"u for helpful discussions. This work is supported by the National Natural Science
Foundation of China (Grant No.~12447134) and the
Postdoctoral Innovation Project of Shandong Province
(Grant No.~SDCX-ZG-202503036).

\appendix

\section{Iyer--Wald derivation of the interior response relation}
\label{app:iyerwald}

We derive the variation of the terminal Kasner coefficients that
enters the source-response relation in Sec.~\ref{sec:landau}.
The calculation compares neighboring stationary black branes,
with the boundary time normalization, horizon gauge, and radial
reference scale held fixed.

For the Lagrangian four-form $\bm L$ in Eq.~\eqref{eq:action},
the variation of the action can be written as
\begin{equation}
 \delta\bm L=\bm E_\Psi\delta\Psi
 +\dd\bm\Theta(\Psi,\delta\Psi),
 \qquad \Psi=(g,A,\phi),
 \label{eq:iwL}
\end{equation}
where $\bm E_\Psi=0$ denotes the field equations and
$\bm\Theta$ is the symplectic potential three-form.
In the normalization $16\pi G_N=1$, its components are
\begin{equation}
 \begin{split}
 \Theta_{abc}=\varepsilon_{dabc}\big[
 g^{df}g^{eg}(\nabla_e\delta g_{fg}
              -\nabla_f\delta g_{eg})-G(\phi)F^{de}\delta A_e
 -\nabla^d\phi\,\delta\phi\big].
 \end{split}
 \label{eq:iwTheta}
\end{equation}

The antisymmetrized field variation of $\bm\Theta$
defines the symplectic current three-form,
\begin{equation}
 \begin{split}
 \bm\omega(\Psi;\delta_1\Psi,\delta_2\Psi)
 ={}&\delta_1\bm\Theta(\Psi,\delta_2\Psi)-\delta_2\bm\Theta(\Psi,\delta_1\Psi).
 \end{split}
 \label{eq:iwomega}
\end{equation}
Its integral over a hypersurface $\Sigma$ is the
presymplectic form
\begin{equation}
 \Omega_\Sigma(\Psi;\delta_1\Psi,\delta_2\Psi)
 =\int_\Sigma
 \bm\omega(\Psi;\delta_1\Psi,\delta_2\Psi).
 \label{eq:iwOmega}
\end{equation}
For a fixed vector field $\xi$, the Hamiltonian variation
generating its diffeomorphism is
\begin{equation}
 \delta H_\xi[\Sigma]
 =\Omega_\Sigma(\Psi;\delta\Psi,\mathcal L_\xi\Psi).
 \label{eq:iwHamiltonian}
\end{equation}

To reduce this bulk integral to a surface integral, consider
the Noether current three-form associated with $\xi$
\cite{Wald1993,IyerWald1994},
\begin{equation}
 \bm J_\xi=\bm\Theta(\Psi,\mathcal L_\xi\Psi)
 -\iota_\xi\bm L.
 \label{eq:iwCurrent}
\end{equation}
On shell, $\bm J_\xi=\dd\bm Q_\xi$, where
$\bm Q_\xi$ is the Noether charge two-form. For
Eq.~\eqref{eq:action},
\begin{equation}
 (Q_\xi)_{ab}=-\varepsilon_{abcd}
 \left[\nabla^c\xi^d
 +\frac12G(\phi)F^{cd}(\xi^eA_e)\right].
 \label{eq:iwNoetherCharge}
\end{equation}
Varying Eq.~\eqref{eq:iwCurrent} along a solution of the
linearized field equations gives the Iyer--Wald identity
\begin{equation}
 \begin{split}
 \bm\omega(\Psi;\delta\Psi,\mathcal L_\xi\Psi)
 &=\dd\bm k_\xi,\quad
 \bm k_\xi=\delta\bm Q_\xi
 -\iota_\xi\bm\Theta(\Psi,\delta\Psi).
 \end{split}
 \label{eq:iwk}
\end{equation}
Let $\Sigma_{12}$ be a hypersurface bounded by the
planar sections $S_{z_1}$ and $S_{z_2}$. By
Eqs.~\eqref{eq:iwHamiltonian} and~\eqref{eq:iwk},
\begin{equation}
 \begin{split}
 \delta H_\xi[\Sigma_{12}]
 &=\int_{\Sigma_{12}}\bm\omega=\int_{S_{z_2}}\bm k_\xi
 -\int_{S_{z_1}}\bm k_\xi.
 \end{split}
 \label{eq:iwStokes}
\end{equation}
The integrals over planar sections will be understood
per unit transverse area. We denote the variation on an
individual radial section by
\begin{equation}
 \delta\calH_\xi(z)=\int_{S_z}\bm k_\xi.
 \label{eq:iwSurfaceH}
\end{equation}
For the stationary black branes, $\xi=\partial_t$ and
$\mathcal L_\xi\Psi=0$. Equation~\eqref{eq:iwStokes}
then gives
\begin{equation}
 \delta\calH_\xi(z_1)=\delta\calH_\xi(z_2).
 \label{eq:iwRadialConservation}
\end{equation}

For the ansatz \eqref{eq:ansatz} with $\xi=\partial_t$,
substituting Eqs.~\eqref{eq:iwTheta} and
\eqref{eq:iwNoetherCharge} into Eq.~\eqref{eq:iwk}
and integrating over a planar section oriented toward
the AdS boundary gives
\begin{equation}
 \delta\calH_\xi(z)
 =e^{-\chi/2}\left[
 -\frac{2}{z^3}\delta f
 +\frac{f\phi'}{z^2}\delta\phi\right]
 -\psi\,\delta\rho.
 \label{eq:iw_var}
\end{equation}
Here $\rho$ is the conserved Maxwell flux in
Eq.~\eqref{eq:maxwell}. We next evaluate this variation
at the RN inner horizon and the terminal Kasner singularity.

At a simple horizon $z=z_h$, varying $f(z_h)=0$
and evaluating \eqref{eq:iw_var} gives
\begin{equation}
 \delta\calH_\xi(z_h)=T_h\,\delta s_h-\psi(z_h)\delta\rho,
 \label{eq:iwHorizonGeneral}
\end{equation}
where $s_h=4\pi/z_h^2$ and
$T_h=-e^{-\chi_h/2}f'(z_h)/(4\pi)$ is the signed
temperature. In the regular gauge $\psi(z_h)=0$ at
the outer horizon, this reduces to
\begin{equation}
 \delta\calH_\xi(z_h)=T\,\delta s.
 \label{eq:iw_h}
\end{equation}

The normal RN solution also possesses a simple
inner horizon $z_->z_h$. Denoting its signed
temperature, entropy density, and potential by
$T_-$, $s_-$, and $\psi_-$, as in
Eq.~\eqref{eq:rn_inner_Ts}, Eq.~\eqref{eq:iwHorizonGeneral}
gives
\begin{equation}
 \delta\calH_\xi(z_-)
 =T_-\,\delta s_--\psi_-\,\delta\rho.
 \label{eq:iw_RN_inner}
\end{equation}
Equating the two radial evaluations yields
\begin{equation}
 T\,\delta s
 =T_-\,\delta s_--\psi_-\,\delta\rho.
 \label{eq:RN_inner_firstlaw}
\end{equation}
This is the inner-horizon first law in the gauge
where the electric potential vanishes at the
outer horizon. Direct substitution of the RN fields
from Eq.~\eqref{eq:RN} into Eq.~\eqref{eq:iw_var}
also gives the $z$-independent expression
\begin{equation}
 \delta\calH_\xi^{\RN}(z)
 =-2\,\delta f_{3,\RN}-\mu\,\delta\rho,
 \label{eq:iwRNcheck}
\end{equation}
where $f_{3,\RN}$ is the coefficient of $z^3$
in the RN metric expansion \eqref{eq:uvmetric}.

For the scalarized black branes, substituting the
terminal asymptotics \eqref{eq:Kasner_phi}--
\eqref{eq:Kasner_psi} into Eq.~\eqref{eq:iw_var}
gives a finite endpoint variation. 
\begin{equation}
 \delta\calH_\xi(\K)
 =\Xi\left(2\,\delta\log f_{\K}
 -v\,\delta\phi_{\K}\right)
 -\psi_{\K}\delta\rho.
 \label{eq:iw_K}
\end{equation}
Comparing this with the outer-horizon result
\eqref{eq:iw_h}, we obtain the Kasner endpoint
first-law identity
\begin{equation}
 T\,\delta s+\psi_{\K}\delta\rho
 =\Xi\left(2\,\delta\log f_{\K}
 -v\,\delta\phi_{\K}\right).
 \label{eq:interior_firstlaw}
\end{equation}
Here $\psi_{\K}$ is the electric potential
difference between the terminal region and the
outer horizon.

The endpoint identity \eqref{eq:interior_firstlaw}
can be related directly to the variation of the
grand potential. Starting from the invariant
$\calK=Ts+\mu\rho$ in Eq.~\eqref{eq:Kthermo},
we find
\begin{equation}
 \begin{split}
 \dd\calK={}&s\,\dd T+\rho\,\dd\mu
 +T\,\dd s+\mu\,\dd\rho\\
 ={}&\Xi\left(2\,\dd\log f_{\K}
 -v\,\dd\phi_{\K}\right)
 +(\mu-\psi_{\K})\,\dd\rho+s\,\dd T+\rho\,\dd\mu.
 \end{split}
 \label{eq:KinteriorVar}
\end{equation}

The grand potential and its boundary first law
are given by Eqs.~\eqref{eq:Fgeneral}
and~\eqref{eq:dF}. Together they imply
\begin{equation}
 \dd\calK=3s\,\dd T+3\rho\,\dd\mu
 +2O\,\dd J-J\,\dd O.
 \label{eq:Kvariation}
\end{equation}
Equating the boundary and Kasner expressions
\eqref{eq:Kvariation} and~\eqref{eq:KinteriorVar},
we arrive at
\begin{equation}
 \begin{split}
 2O\,\dd J-J\,\dd O={}&
 \Xi\left(2\,\dd\log f_{\K}
 -v\,\dd\phi_{\K}\right)\\
 &+(\mu-\psi_{\K})\,\dd\rho
 -2s\,\dd T-2\rho\,\dd\mu.
 \end{split}
 \label{eq:master}
\end{equation}
At fixed $(T,\mu)$, this reduces to
Eq.~\eqref{eq:masterfixed}. The variation follows
a family of regular stationary solutions with
variable boundary scalar source $J$, providing the
relation between the Kasner coefficients and the
boundary Landau response used in Sec.~\ref{sec:landau}.

\bibliographystyle{JHEP}
\bibliography{references}
\end{document}